\documentclass[twocolumn,aps,prd,superscriptaddress,preprintnumbers,nofootinbib,floatfix]{revtex4-1}

\usepackage{amsmath}
\usepackage{graphicx}
\usepackage{hyperref}
\usepackage{breakurl}

\usepackage{xcolor}
\definecolor{red}{rgb}{1, 0, 0}

\newcommand{\w}{\omega}
\newcommand{\nn}{\nonumber}
\newcommand{\beq}{\begin{equation}}
\newcommand{\eeq}{\end{equation}}
\newcommand{\beqa}{\begin{eqnarray}}
\newcommand{\eeqa}{\end{eqnarray}}

\newcommand{\GeV}{\rm GeV}
\newcommand{\MeV}{\rm MeV}

\def\ov{\overline}

\def\lqcd{\Lambda_{\rm QCD}}

\def\d{{\rm d}}

\newcommand{\Bbar}{\,\overline{\!B}{}}
\newcommand{\Dbar}{\,\overline{\!D}{}}
\newcommand{\Kbar}{\,\overline{\!K}{}}
\def\B0bar{\Bbar{}^0}
\def\D0bar{\Dbar{}^0}
\def\K0bar{\Kbar{}^0}

\newcommand{\hqs}{\ensuremath{\hat q^2}}
\newcommand{\hq}{\ensuremath{\hat q}}
\def\rt{\ensuremath{\rho_\tau}}

\begin{document}

\preprint{\vbox{
\hbox{DESY 14-095}
}}

\title{\boldmath Precise predictions for $B\to X_c\tau\bar\nu$ decay distributions}

\author{Zoltan Ligeti}
\affiliation{Ernest Orlando Lawrence Berkeley National Laboratory,
University of California, Berkeley, CA 94720}

\author{Frank J.\ Tackmann}
\affiliation{Theory Group, Deutsches Elektronen-Synchrotron (DESY), D-22607 Hamburg, Germany}

\begin{abstract}

We derive precise standard model predictions for the dilepton invariant mass and
the $\tau$ energy distributions in inclusive $B\to X_c\tau\bar\nu$ decay.  We
include $\lqcd^2/m_b^2$ and $\alpha_s$ corrections using the $1S$ short-distance
mass scheme, and estimate shape function effects near maximal $\tau$ energy.
These results can improve the sensitivity of $b\to c\tau\bar\nu$ related
observables to beyond standard model physics.

\end{abstract}

\maketitle

\section{Introduction}
\label{sec:intro}

Recently, $B$ decays mediated by $b\to c\tau\bar\nu$ transitions have received
renewed attention due to improved measurements of the $\bar{B} \to
D\tau\bar{\nu}$ and $\bar{B}\to D^*\tau\bar{\nu}$ decay
rates~\cite{Lees:2012xj}, consistent with earlier published~\cite{Bozek:2010xy,
Aubert:2007dsa} and preliminary~\cite{FPCPtalk} results. Considering the ratios
($\ell = e,\mu$)
\begin{equation}\label{Rdef}
  R(X) = \frac{\mathcal{B}(B \to X \tau\bar\nu)}
              {\mathcal{B}(B \to X \ell \bar\nu)}\,,
\end{equation}
the combination of the BaBar results
\begin{equation}
  \label{recentdata}
  R(D^*) = 0.332 \pm 0.030 \,, \qquad 
  R(D)   = 0.440 \pm 0.072 \,,
\end{equation}
gives a more than $3\sigma$ deviation~\cite{Lees:2012xj} from the standard model
(SM), which could indicate new physics that couples non-universally to leptons,
due to $m_\tau \gg m_{e,\mu}$.  The isospin-constrained fit for the branching ratios
yields~\cite{Lees:2012xj}
\beq\label{sumoftwo}
  \mathcal{B}(\bar{B} \to D^*\tau\bar{\nu}) + 
  \mathcal{B}(\bar{B} \to D \tau\bar{\nu}) = (2.78 \pm 0.25)\, \%\,.
\eeq
(This average applies for $B^-$ decay~\cite{Lees:2012xj}; recall the
lifetime difference of $B^\pm$ and $B^0$.)

A recent update of the SM prediction for $R(X_c)$, the ratio for inclusive decay
rates, yields~\cite{inprep}
\beq\label{RXc}
R(X_c) = 0.223 \pm 0.005\,,
\eeq
which, combined with the world average, $\mathcal{B}(B^-
\to X_c e \bar{\nu}) = (10.92 \pm 0.16)\%$~\cite{Bernlochner:2012bc,
Amhis:2012bh}, yields~\cite{inprep}
\beq\label{rate}
\mathcal{B}(B^- \to X_c\tau\bar\nu) = (2.42 \pm 0.06)\%\,.
\eeq
This prediction is rather precise, thus the inclusive measurement can provide
information complementary to those from the exclusive modes.

The results in Eq.~(\ref{sumoftwo}) are in some tension with the LEP average of
the rate of an admixture of $b$-flavored hadrons to decay to $\tau$
leptons~\cite{PDG}
\beq
\mathcal{B}(b \to X\tau^+\nu) = (2.41 \pm 0.23)\%\,.
\eeq
This rate has not been measured since the LEP experiments.  Neither are
theoretical predictions available for $B \to X\tau\bar\nu$ decay distributions
using a well-defined short-distance quark mass scheme.  Such predictions are
necessary to provide the best theoretical inputs for future experimental
measurements.  Measuring the inclusive rate should be possible using the
existing $B$ factory data, and especially using the future Belle~II data
set~\cite{Phill}.

In the future, the uncertainties of the individual $\bar{B}\to
D^{(*)}\tau\bar{\nu}$ branching ratios are expected to be reduced to about 2\%
by Belle~II~\cite{Aushev:2010bq}, while the uncertainties of the ratios in
Eq.~(\ref{recentdata}) may become even smaller.  Clearly, both inclusive and
exclusive measurements should be pursued.

\section{The OPE results}

Inclusive semileptonic $B$ decay rates can be computed model independently in an
operator product expansion (OPE) in terms of local heavy-quark operators
(for a review, see Ref.~\cite{Manohar:2000dt}).  The
leading order reproduces the free-quark decay result, and perturbative and
nonperturbative corrections can be systematically incorporated.

The triple differential distribution has been derived, including the leading
nonperturbative corrections of order $\lqcd^2/m_b^2$, in
Refs.~\cite{Falk:1994gw, Balk:1993sz, Koyrakh:1993pq}. 
We use the dimensionless kinematic variables
\begin{equation}
\hqs = \frac{q^2}{m_b^2}\,,\qquad
  v\cdot \hat q = \frac{v \cdot q}{m_b}
\,,\qquad
y = {2E_\tau\over m_b}
\,,\end{equation}
where $q = p_\tau + p_\nu$ is the dilepton momentum, $v$ is the four-momentum of
the $B$ meson [$(1, \vec 0)$ in the $B$ rest frame], and $E_\tau = v\cdot p_\tau$
is the $\tau$ energy measured in the $B$-meson rest frame.
The mass parameters are defined as
\beq
\rt = \frac{m_\tau^2}{m_b^2}
\,,\qquad
x_\tau = \frac{m_\tau^2}{q^2} = \frac{\rt}{\hqs}
\,,\qquad
\rho = {m_c^2\over m_b^2}\,.
\eeq
It is convenient to define
\begin{equation}
y_\pm = \frac12 \Big(y \pm  \sqrt{y^2-4\rt}\Big) 
.\end{equation}
Then $y_+ y_- = \rho_\tau$, and $\{y_+,\, y_-\} \to \{y,\, 0\}$ as $m_\tau\to
0$.

The triple differential decay rate in the $B$ rest frame is
\begin{align} \label{triplediff}
&\frac{1}{\Gamma_0} \frac{\d\Gamma}{\d\hqs\, \d y\, \d v\cdot\hq}
\nn \\ & \quad
= 24\, \theta\bigl[(2v\cdot \hat q - y_+) y_+ - \hqs\bigr]\, 
  \theta\bigl[\hqs - (2v\cdot \hat q - y_-) y_-\bigr]
\nn \\ & \qquad \times
\biggl\{2 (\hqs - \rt) \hat W_1 + \bigl[y(2 v\cdot\hat q - y) - \hqs + \rt\bigr]\,\hat W_2
\nn \\ & \qquad\quad
+ 2 \bigl[\hqs (y - v\cdot \hat q) - \rt v\cdot \hat q\bigr]\,\hat W_3
\nn \\ & \qquad\quad
+ \rt (\hqs - \rt)\, \hat W_4 + 2\rt (2 v\cdot\hat q - y)\, \hat W_5 \biggr\}
,\end{align}
where
\beq
\Gamma_0 = {\lvert V_{cb} \rvert^2\, G_F^2\, m_b^5 \over 192\pi^3}
\eeq
is the tree-level free-quark decay rate.
The $\hat W_i$ are the structure functions of the hadronic
tensor~\cite{Manohar:1993qn, Falk:1994gw}, which in the local OPE to
$\lqcd^2/m_b^2$ contain $\delta$, $\delta'$, and $\delta''$ functions of $(1+\hqs
- 2v\cdot\hat q -\rho)$.

In the literature, only the $E_\tau$ spectrum and the total decay rate have been
computed including $\lqcd^2/m_b^2$ corrections~\cite{Falk:1994gw, Balk:1993sz,
Koyrakh:1993pq} (as well as the $\tau$ polarization~\cite{Falk:1994gw}).   These
corrections reduce the $B\to X_c\tau\bar\nu$ rate by about 7\,--\,8\%, where
about 90\% of this reduction is due to the terms proportional to $\lambda_2$.  

In this paper, we also derive the order $\lqcd^2/m_b^2$ corrections for the
$q^2$ spectrum, as it is expected to be useful for the experimental
analysis~\cite{Phill}.  While the perturbative corrections were known in the
literature in the pole mass scheme, only the total rate was calculated in a
short-distance mass scheme in the past.  We present results for the first time
for the $q^2$ and $E_\tau$ spectra in a well-defined short-distance mass scheme.
In addition, we pay special attention to the uncertainties in the end point
regions of these spectra, where the local OPE breaks down.

\subsection{Phase space limits}
\label{subsec:phasespace}

\begin{figure}[tb]
\includegraphics[width=\columnwidth]{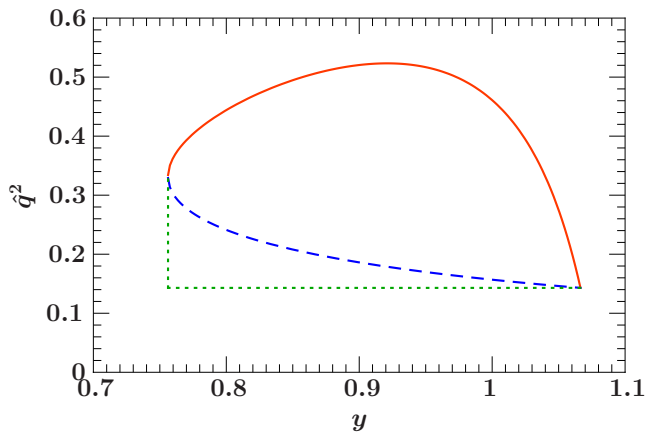}
\caption{The $b\to c\tau\bar\nu$ Dalitz plot for free quark decay. The solid
(orange) boundary comes from the first $\theta$ function in
Eq.~(\ref{triplediff}), the dashed (blue) boundary from the second one.}
\label{dalitz}
\end{figure}

A complication in the massive lepton case is the appearance of the second
$\theta$ function in Eq.~\eqref{triplediff}, which sets a nontrivial lower limit
on $\hqs$ (which in the $m_\tau \to0$ limit reduces to $\hqs > 0$). Solving
the $\theta$ functions for the limits on $y$ for fixed $\hqs$ and $v\cdot\hat
q$, we have
\begin{equation} \label{eq:ylimits}
\hat q_- + x_\tau\, \hat q_+ \leq y \leq \hat q_+ + x_\tau\, \hat q_-
\,,\end{equation}
where
\begin{equation}
\hat q_\pm = v\cdot \hat q \pm \sqrt{(v\cdot \hat q)^2 - \hqs}
\,.\end{equation}
Substituting the parton level result for $v\cdot \hat q = (1+\hqs- \rho)/2$ then
gives partonic phase space in the $\hqs-y$ plane at tree level. The limits on
$\hqs$ for fixed $y$ are
\begin{equation} \label{eq:q2limits}
y_- \Bigl(1 - \frac{\rho}{1 - y_-} \Bigr) \leq \hqs \leq y_+ \Bigl(1 - \frac{\rho}{1 - y_+} \Bigr)
\,.\end{equation}
This is shown in Fig.~\ref{dalitz}, where we used $\rho=(1.3/4.7)^2$ and $\rt =
(1.777/4.7)^2$ for illustration. The solid (orange) boundary comes from the
first $\theta$ function in Eq.~\eqref{triplediff}, and the dashed (blue)
boundary comes from the second one.

Note that the limits for $y$ are determined by the different
$\theta$ functions for values of $q^2$ above and below
\begin{equation}
\hqs_0 = \sqrt{\rho_\tau}\, \biggl(1 - \frac\rho{1-\sqrt{\rho_\tau}} \biggr)
\,.\end{equation}
A similar situation occurs in the
calculation of the ${\cal O}(\alpha_s)$ correction to $\d\Gamma/\d y\,
\d\hqs$~\cite{Jezabek:1996db}, but was not encountered in calculating
${\cal O}(\lqcd^2/m_b^2)$ corrections before.

Beyond tree level, the lower limit of the $\d\hqs$ integration and the lower
limit of $\d y$ integration for $\hqs < \hqs_0$ (blue dashed curve in
Fig.~\ref{dalitz}) gets replaced by $\hqs > \rt$ and $y > 2\sqrt{\rt}$, which is
shown by the dotted (green) lines.

Integrating over $\hqs$, the limits of the $y$ spectrum are
\begin{equation} \label{ylim}
2\sqrt\rt < y < 1 + \rt - \rho
\,.\end{equation}
Integrating over $y$, the overall limits of the $\hqs$ spectrum are
\beq\label{hqslim}
\rt < \hqs < (1 - \sqrt\rho)^2\,.
\eeq
The above are the partonic phase space limits relevant to the OPE result. For
the hadronic phase space limits, $m_b$ is replaced by $m_B$ and $\rho$ is
replaced by $m_D^2/m_B^2$.

\subsection{\boldmath The $q^2$ spectrum}
\label{subsec:q2}

Since the hadronic structure functions $\hat W_i$ are functions of $\hq^2$ and
$\hat v\cdot q$ only, it is easiest to first integrate the triple differential
spectrum in Eq.~\eqref{triplediff} over the lepton energy with the limits given
in Eq.~\eqref{eq:ylimits}. Doing so, we obtain for the double differential
spectrum
\begin{align}
\frac{1}{\Gamma_0} \frac{\d\Gamma}{\d\hqs\, \d v\cdot\hq}
&= 96\, (1 - x_\tau)^2 \sqrt{(v\cdot\hat q)^2 - \hqs} \\ 
& \quad\times \Bigl\{ \hqs\, \hat W_1 
+ \frac{1}{3} \bigl[(v\cdot\hat q)^2 - \hqs\bigr](1 + 2x_\tau)\, \hat W_2
\nn\\ & \qquad
+ \frac{\rt}{2}\, \bigl(\hat W_2 + \hqs \hat W_4 + 2v\cdot \hat q \hat W_5\bigr) \Bigr\}
\,.\nn\end{align}
Substituting the OPE results for the $\hat W_i$, we obtain for the $q^2$
spectrum
\begin{align}\label{q2spec}
\frac{1}{\Gamma_0} \frac{\d\Gamma}{\d\hqs}
&= 2 (1 - x_\tau)^2 \sqrt{P^2 - 4\rho}\,
\biggl\{\! \biggl(1 + \frac{\lambda_1 + 15\lambda_2}{2m_b^2}\biggr)
\nn\\ & \quad\times
\Bigl[3 \hqs P (1 + x_\tau) + (P^2 - 4\rho) (1 + 2 x_\tau) \Bigr]
\\ & \quad+
\frac{6\lambda_2}{m_b^2} \bigg[ (P - 2) (1 + 2 x_\tau) + \hqs (4 + 5 x_\tau)
\nn \\ & \quad
  + \hqs\frac{2 (2\hqs \!+\! P\!-\!2) (2 + x_\tau) + 3 \hqs P (1 + x_\tau)}{P^2 - 4\rho}
\biggr] \!\biggr\}
,\nn\end{align}
where we defined $P = 1  - \hqs + \rho$, and we have suppressed the $\theta$
functions expressing the $\hqs$ limits given in Eq.~\eqref{hqslim}.
Integrating over $\hqs$ we reproduce the total rate given in Ref.~\cite{Falk:1994gw}.

As we will see in Sec.~\ref{sec:numerics} below, the order $\lqcd^2/m_b^2$
corrections reduce the $B\to X_c\tau\bar\nu$ rate mainly at higher values of
$\hqs$, dominated by the terms proportional to $\lambda_2$.  Near maximal
$\hqs$, the $\lambda_2$ terms behave as $(\hqs_{\rm max} - \hqs)^{-1/2}$, and
the differential rate becomes negative.  This indicates a breakdown of the OPE;
in this region of phase space, the hadronic final state is constrained to be in
the resonance region, and the OPE cannot describe the spectrum point by point. 
Thus, integration over some region of $\Delta\hqs$ is necessary near maximal
$\hqs$ to obtain a reliable result.  The form of Eq.~(\ref{q2spec}) makes it
clear that this effect is not related to the $b$ quark distribution function in
the $B$ meson, the so-called shape function (which is neither relevant for the
high $q^2$ region in $B\to X_u\ell\bar\nu$~\cite{Bauer:2000xf}).  Note also that
the difference of the upper limit of $q^2$ at lowest order in the OPE and at the
hadronic level is suppressed by $\lqcd^2$.

\subsection{The \boldmath $\tau$ energy spectrum}
\label{subsec:Etau}

To obtain the $E_\tau$ spectrum, we substitute the OPE results for the $\hat
W_i$ in the triple differential rate in Eq.~(\ref{triplediff}).  The integration
over $v\cdot \hat q$ is performed using the $\delta^{(n)}(1 + \hqs - 2v\cdot\hat
q - \rho)$ contained in the $\hat W_i$. Next, we integrate over $\hqs$ with the
integration limits in Eq.~(\ref{eq:q2limits}). At leading order we obtain
\begin{align} \label{eq:dGdyLO}
\frac{1}{\Gamma_0} \frac{\d\Gamma}{\d y}
&= 2 \sqrt{y^2 - 4\rt}\, \theta(y - 2\rt)\,\theta(1 - R)\, (1 - R)^2 \\
& \quad\times \Bigl[
y \rho\, \frac{1-R}{R} + (1 + 2R)(y - 2\rt)(2-y) \Bigr]\nn
\,,\end{align}
where
\begin{equation}
R = \frac{\rho}{(1 - y_+)(1 - y_-)} = \frac{\rho}{1 - y + \rt}
\,,\end{equation}
and for the $\lqcd^2/m_b^2$ corrections we reproduce the results in
Refs.~\cite{Falk:1994gw, Balk:1993sz, Koyrakh:1993pq}. The two $\theta$
functions in Eq.~\eqref{eq:dGdyLO} correspond to the limits on $y$ in
Eq.~\eqref{ylim}.

As for large values of $\hqs$, the OPE also breaks down for large values of
$y$.  Contrary to the end point of the $q^2$ spectrum, the $E_\tau$ end point does
differ by an amount of order $\lqcd$ between the partonic and hadronic phase
space limits.  If one treats $m_c \sim {\cal O}(\sqrt{m_b\,\lqcd})$, or equivalently
$\rho \sim \lqcd/m_b$, then the problematic terms in the OPE that are
enhanced near the end point can be resummed, replacing the usual OPE by
an expansion in terms of nonlocal light-cone operators, whose matrix elements
yield nonperturbative $B$-meson distribution functions (shape functions).  [Such
effects would formally be subleading if one treats $m_c^2/m_b^2 \sim {\cal O}(1)$.]  At the lowest
order description of the end point region, ${\cal O}[(\lqcd/m_b)^0]$, a single
shape function appears. This is well known for $B\to X\ell\nu$
decays~\cite{Neubert:1993ch, Falk:1993vb, Bigi:1993ex, Mannel:1994pm}.  When carried out
appropriately, the shape function OPE can be rendered valid away from the
end point region as well, such that it smoothly recovers the local OPE
result~\cite{Mannel:2004as, Tackmann:2005ub}. For $b\to c$ transitions, this is
possible if the OPE is directly performed for the lepton energy
spectrum~\cite{Mannel:2004as}. Following Ref.~\cite{Mannel:2004as} and including
the $\tau$ mass, we obtain at leading order
\begin{align} \label{eq:dGdySF}
\frac{1}{\Gamma_0} \frac{\d\Gamma}{\d y}
&= 2 \sqrt{y^2 - 4\rt}\, \int\!\d\hat \w\, m_b\, F(m_b \hat\w + m_B - m_b)
\nn\\  & \quad\times
\theta(y - 2\rt) \theta(1 - R_\w)\, (1 - R_\w)^2\,
\Bigl\{ y\rho\, \frac{1-R_\omega}{R_\omega}
\nn\\
&\qquad + (1 + 2R_\w) \bigl[y - \hat\w y_- - 2\rt\bigr] (2 - y - \hat\w) \Bigr\}
.\end{align}
where
\begin{equation}
R_\w = \frac{\rho}{(1 - y_+ - \hat\w)(1 - y_-)}
\,,\end{equation}
and the leading shape function, $F(k)$, in Eq.~\eqref{eq:dGdySF} is defined with
the same conventions as in Refs.~\cite{Tackmann:2005ub, Ligeti:2008ac}.

For $B\to X_c\tau\nu$, the end point region of the lepton energy spectrum is
given by $1 - y_+ \sim \lqcd/m_b$. The result in Eq.~\eqref{eq:dGdySF} arises
from replacing $1-y_+ \to 1-y_+-\hat k_+$ in the local OPE. (Some overall
factors that arise from the leptonic phase space are unaffected.) For small
$y_+$, corresponding to small $E_\tau$, one can expand
\begin{equation}
m_b F(m_b \hat\w + m_B - m_b) = \delta(\hat\w) + \ldots
\,,\end{equation}
which recovers the leading-order result in Eq.~\eqref{eq:dGdyLO}.
In principle, all $\lqcd^2/m_b^2$ corrections in the local OPE at small $y_+$
can be recovered from the shape function expansion, which would require one to
carry it out to the same higher order~\cite{Tackmann:2005ub}.

\subsection{\boldmath The $1S$ mass scheme and perturbative corrections}

It is well known that the pole mass of a heavy quark is not well defined beyond
perturbation theory.  This manifests itself, for example, in poorly behaved
perturbation series.  In this paper, we use the $1S$ mass
scheme~\cite{Hoang:1998ng, Hoang:1998hm, Hoang:1999zc}.
Including both the $c$ quark and $\tau$ lepton mass effects, the corrections to
free-quark decay for the total rate were computed to ${\cal
O}(\alpha_s)$~\cite{Hokim:1983yt}, ${\cal O}(\alpha_s^2
\beta_0)$~\cite{Lu:1996jh}, and ${\cal O}(\alpha_s^2)$~\cite{Biswas:2009rb}. 
The ${\cal O}(\alpha_s)$ result~\cite{Hokim:1983yt} was already used in the
numerical prediction for the rate 20 years ago~\cite{Falk:1994gw}, and the
${\cal O}(\alpha_s^2\beta_0)$ result~\cite{Lu:1996jh} could be used to show that
the perturbation series in the $1S$ scheme,
$1 - 0.070\epsilon - 0.016 \epsilon^2_{\rm BLM}$~\cite{Hoang:1998hm},
is much better behaved than that in the pole scheme, $1-0.097\epsilon - 0.064
\epsilon^2_{\rm BLM}$.  (Here powers of $\epsilon=1$ indicate the order in the
$1S$ expansion, and $\epsilon^2_{\rm BLM}$ corresponds to the lowest order term
proportional to $\beta_0=11-2n_f/3$, the first coefficient in the QCD $\beta$
function.)  This improvement in the perturbation series is essential to obtain
the precise predictions in Eqs.~(\ref{RXc}) and (\ref{rate}).

The ${\cal O}(\alpha_s)$ correction to $\d\Gamma/\d \hqs$ was calculated
analytically in Ref.~\cite{Czarnecki:1994bn}, while the corrections to the
lepton energy spectrum can be obtained by integrating $\d^2\Gamma/\d y\,\d \hqs$
calculated in Ref.~\cite{Jezabek:1996db}.  In particular, the fractional
corrections at order $\alpha_s$ to both $\d\Gamma/\d \hqs $ and $\d\Gamma/\d y$
are remarkably independent of $\hqs$ and $y$, and so have very little effect on
the shape of the spectra except very close to their end points.

\section{\boldmath Numerical results}
\label{sec:numerics}

\begin{figure*}[tb]
\centerline{\includegraphics[height=5.8cm]{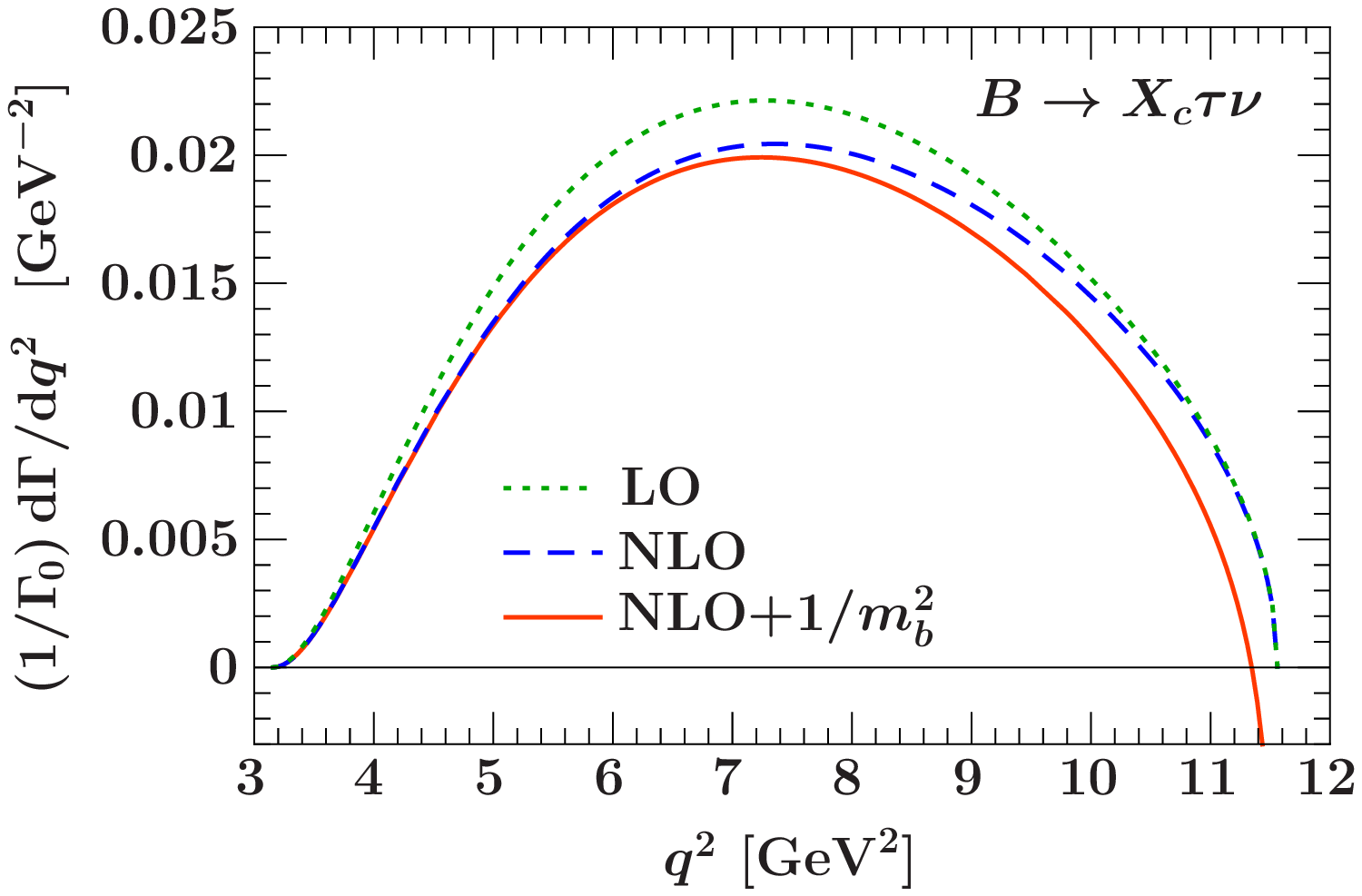} \hfill
\includegraphics[height=5.8cm]{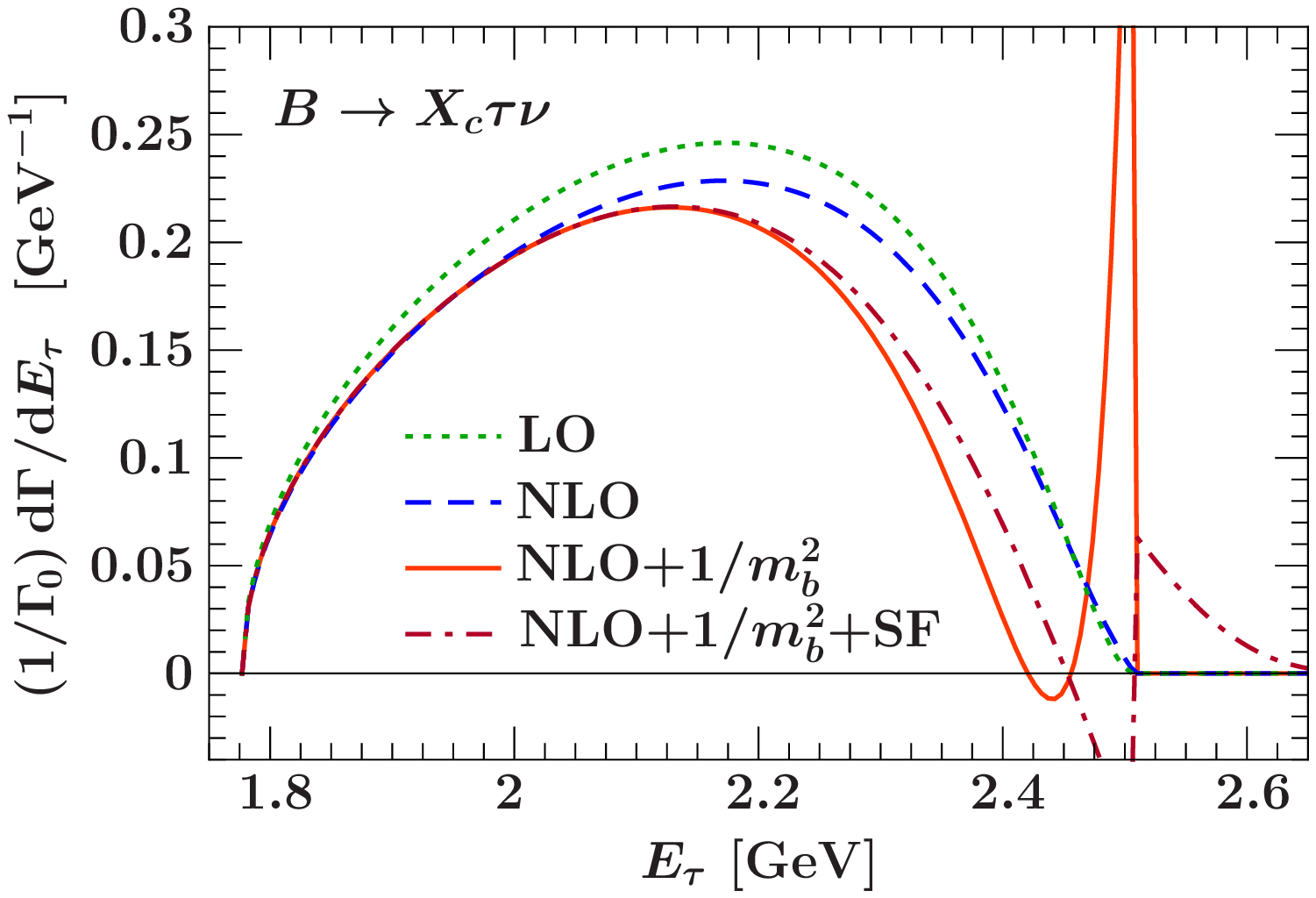}}
\caption{The OPE predictions for the $\d\Gamma/\d q^2$ (left) and $\d\Gamma/\d
E_\tau$ (right) in $B\to X_c\tau\bar\nu$.  The dotted (green) curves show the
free-quark decay result, the dashed (blue) curves include ${\cal O}(\alpha_s)$
corrections, and the solid (orange) curves include both $\alpha_s$ and
$\lqcd^2/m_b^2$ corrections. For $\d\Gamma/\d E_\tau$ the dot-dashed (dark red)
curve combines ${\cal O}(\alpha_s,\, \lqcd^2/m_b^2)$ with the leading-order
shape function result.}
\label{fig:spectra}
\end{figure*}

\begin{figure*}[tb]
\centerline{\includegraphics[width=\columnwidth]{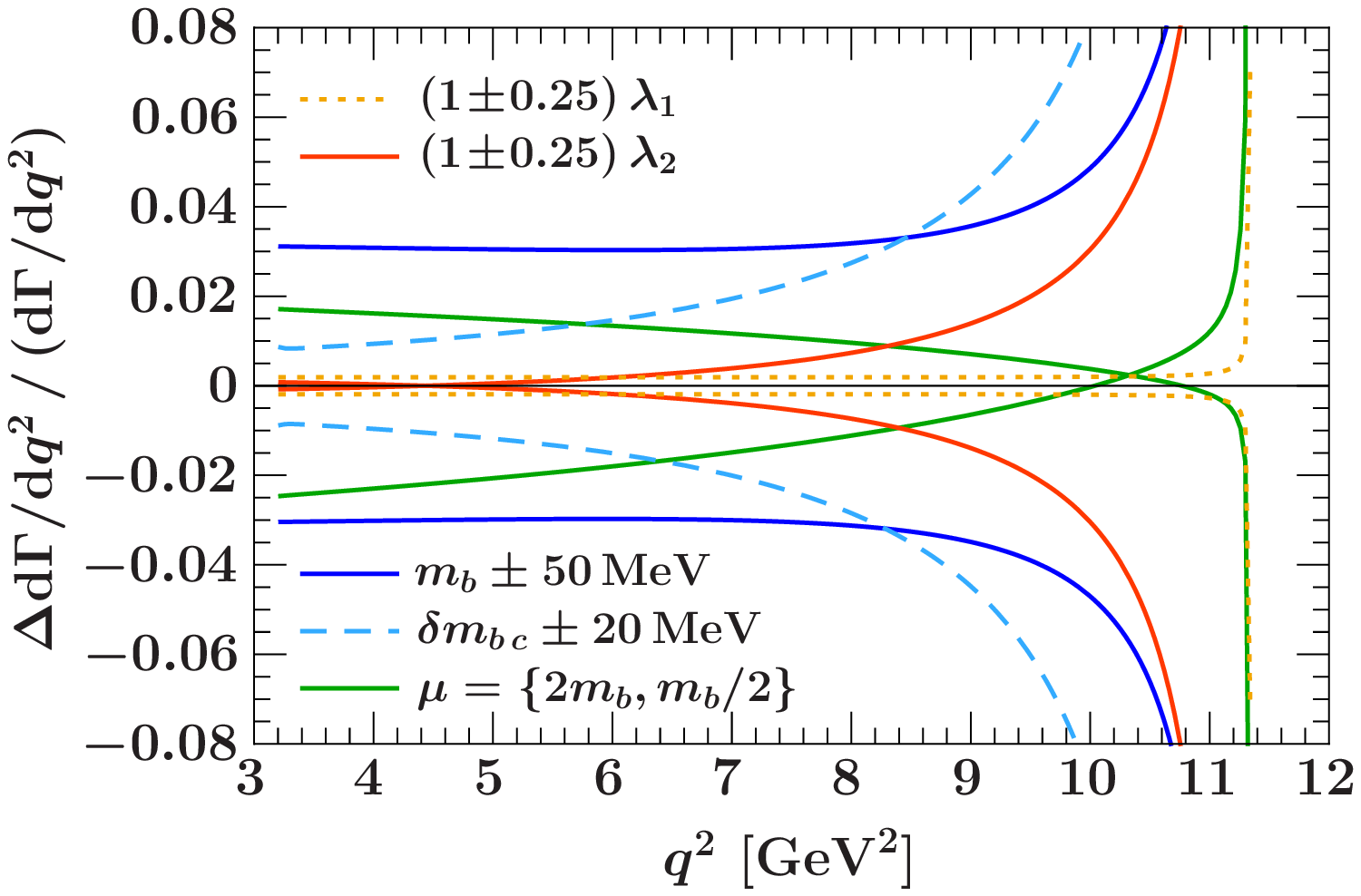} \hfill
\includegraphics[width=\columnwidth]{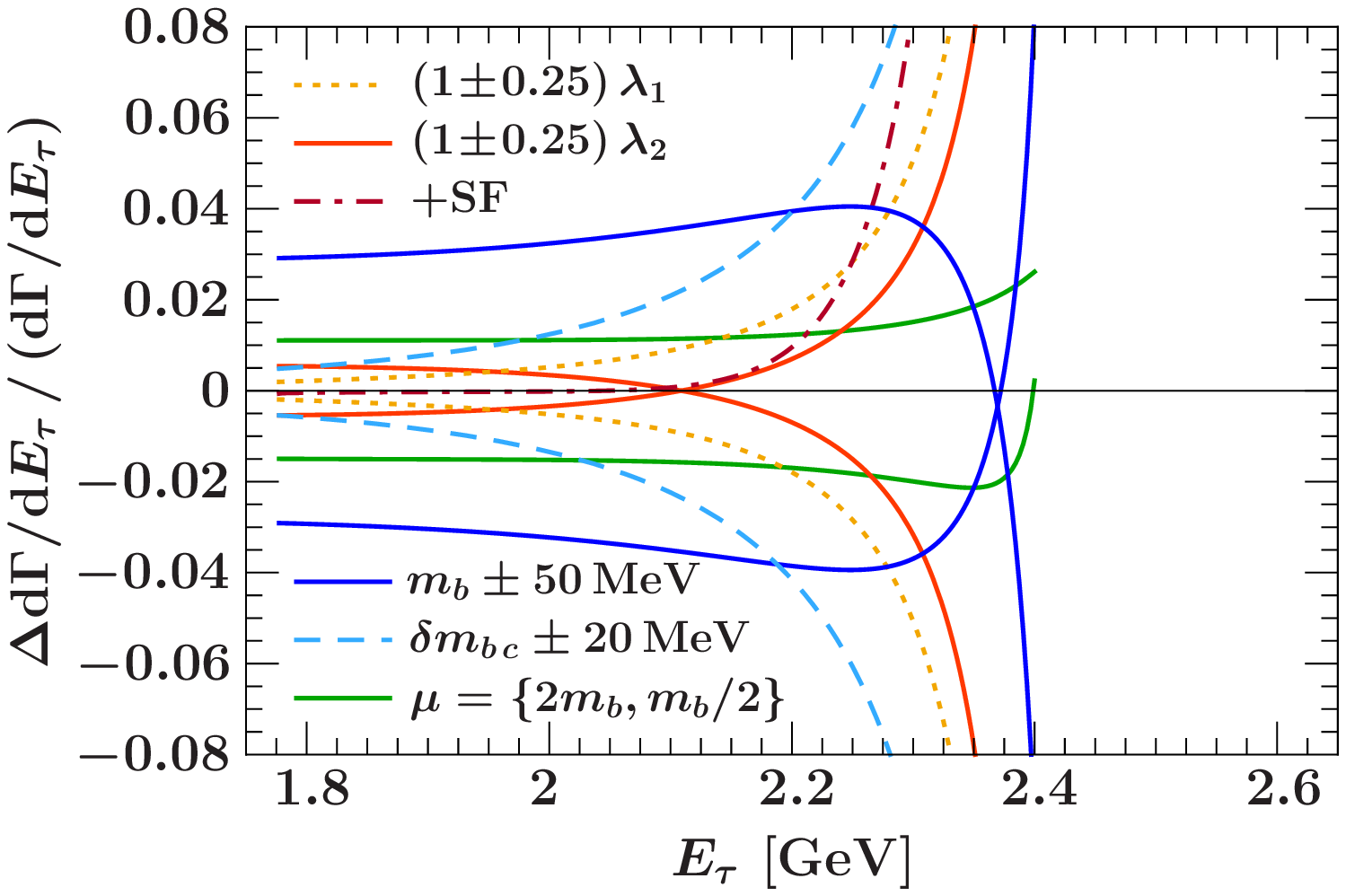}}
\caption{The fractional uncertainties in the OPE predictions for $\d\Gamma/\d
q^2$ (left) and $\d\Gamma/\d E_\tau$ (right). The solid blue curves show the
effect of the variation of $m_b^{1S}$ by $\pm50\,\MeV$ (keeping $\delta m_{bc}$
fixed), the dashed light blue curves show the variation of $\delta m_{bc}$ by
$\pm20\,\MeV$, the solid green curves show the $\mu$ variation between $m_b/2$
and $2m_b$, and the solid red (dotted light orange) curves show the variation of
the coefficient of $\lambda_2$ ($\lambda_1$) by $\pm25\%$. The dot-dashed (dark red)
curve shows the relative correction from including the leading shape function.}
\label{fig:spectvar}
\end{figure*}

Hereafter we revert to dimensionful kinematic variables, $E_\tau$ and $q^2$
(i.e., no longer rescale them by powers of $m_b$).  The phase space limits for
the $\hqs$ and $y$ distributions are given in Eqs.~\eqref{hqslim} and \eqref{ylim}.
Restoring the dimensions of the variables,
\beqa
m_\tau &<& E_\tau < \frac{m_b^2-m_c^2+m_\tau^2}{2m_b}\,, \nn\\
m_\tau^2 &<& q^2 < (m_b-m_c)^2\,.
\eeqa
One can immediately see, writing 
\beq
m_{b,c} = m_{B,D} - \bar\Lambda + {\cal O}(\lqcd^2/m_{b,c}^2)\,,
\eeq
that the difference of the upper limit of $q^2$ at lowest order in the OPE,
$(m_b-m_c)^2$, and at the hadronic level, $(m_B-m_D)^2$, is suppressed by
$\lqcd^2$.  However, the lepton energy end point does receive an ${\cal
O}(\lqcd)$ correction, although only about 100\,MeV (it is $\sim 300$\,MeV for
$B\to X_u e\bar\nu$).  As explained above, treating $m_c^2 / (m_b\lqcd) \sim
{\cal O}(1)$ or $m_c^2/m_b^2 \sim {\cal O}(1)$ affects whether the shape
function is formally relevant to describe the $E_\tau$ end point region. We use
Eq.~\eqref{eq:dGdySF} to determine beyond which value of $E_\tau$ the shape
function becomes important and the local OPE result cannot be trusted anymore. A
more detailed analysis for $B\to X_u\tau\nu$ will be given
elsewhere~\cite{prep}.

\begin{figure*}[tb]
\centerline{\includegraphics[width=\columnwidth]{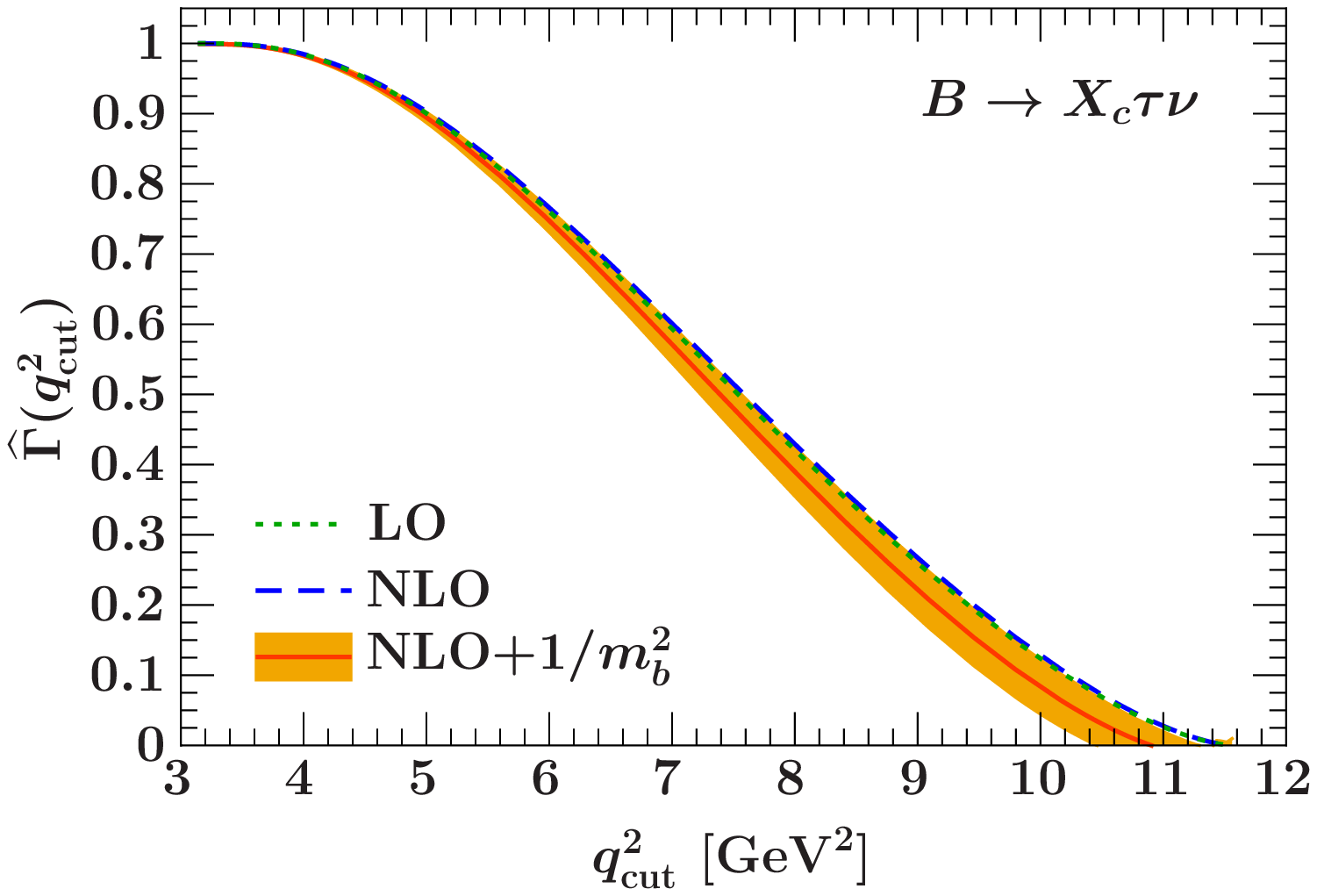} \hfill
\includegraphics[width=\columnwidth]{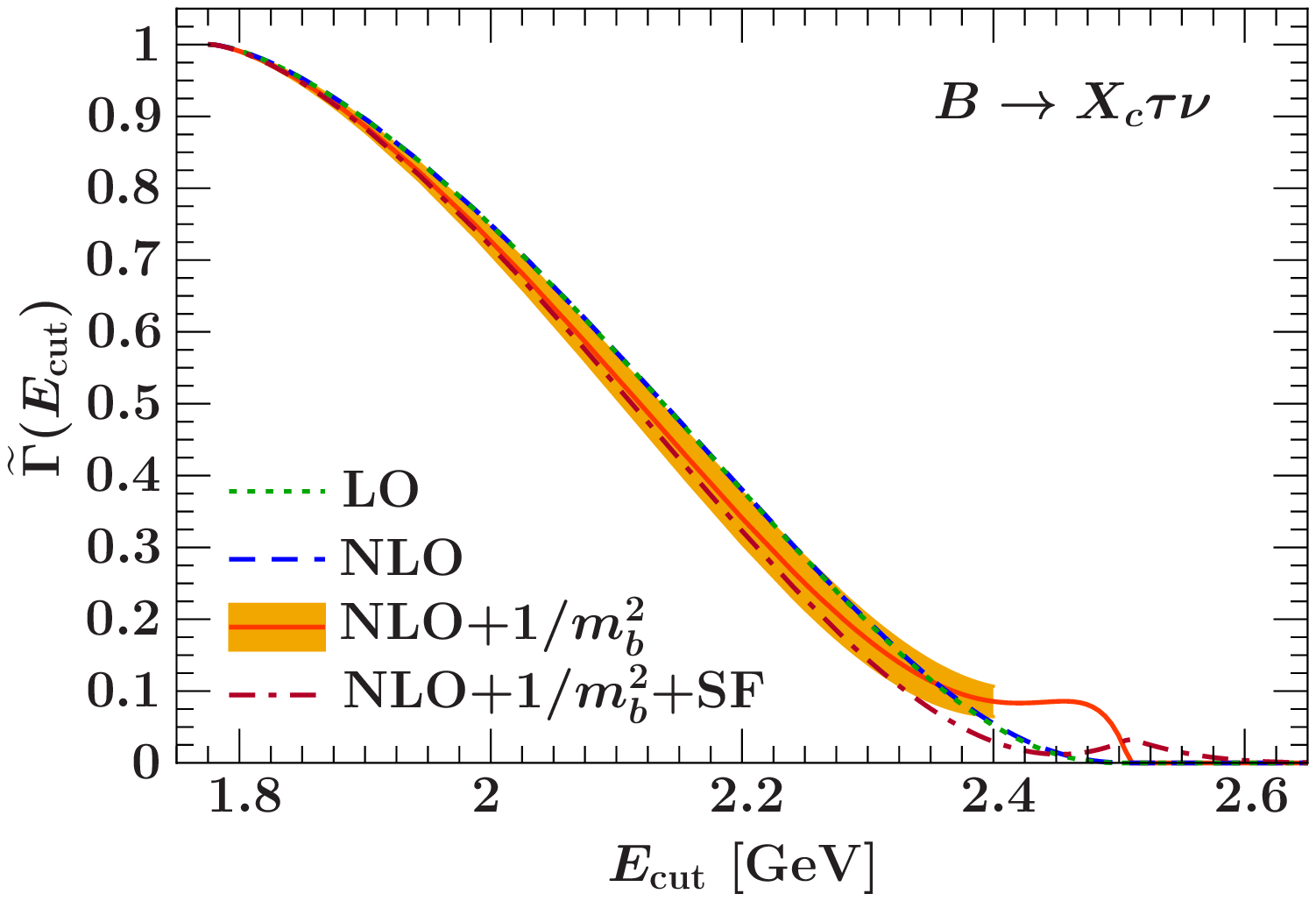}}
\caption{The OPE predictions for the fraction of events above a certain cut in
$\d\Gamma/\d q^2$ (left) and $\d\Gamma/\d E_\tau$ (right) in $B\to
X_c\tau\bar\nu$. The meaning of the curves is the same as in Fig.~\ref{fig:spectra}.
The shaded band shows the total uncertainties in the full result.}
\label{fig:cumulant}
\end{figure*}

\begin{table}[b]
\tabcolsep 8pt
\begin{tabular}{c|ccc}
\hline\hline
Parameter  &  Central value & Variation  &  $\Delta\Gamma_{\rm total}$ \\ \hline
$m_b^{1S}$  &  $4.71\,\GeV$  &  $\pm50\,\MeV$  &  $\pm5.3\,\%$ \\
$\delta m_{bc}$  &  $3.40\,\GeV$  &  $\pm20\,\MeV$  &  $\pm4.4\,\%$ \\
$\lambda_1$  &  $-0.30\,\GeV^2$  &  $\pm25\%$  &  $\pm0.2\,\%$ \\
$\lambda_2$  &  $0.12\,\GeV^2$  &  $\pm25\%$  &  $\pm2.0\,\%$ \\
$\alpha_s$  &  $0.218$  &  ${}^{+0.065}_{-0.040}$  &  $\pm1.1\,\%$ \\
\hline\hline
\end{tabular}
\caption{Central values of input parameters, their variations, and the resulting
uncertainties in the total rate prediction.}
\label{parameters}
\end{table}

The numerical inputs we use are summarized in Table~\ref{parameters}. For the
leading order shape function, we use the fit result from
Ref.~\cite{Bernlochner:2013bta}, and for consistency we also take the central
value for $m_b^{1S}$ from there, which is consistent with the fit results in the
$1S$ scheme in Refs.~\cite{Bauer:2004ve, Amhis:2012bh}, with a conservative
error of $\pm 50\,\MeV$.

In the $1S_{\rm EXP}$ scheme in Ref.~\cite{Bauer:2004ve}, one relates $m_b-m_c$
using heavy-quark effective theory (HQET) to a linear combination of the spin averaged hadron mass difference,
$\ov m_B-\ov m_D$, $\lambda_1$, and dimension-6 HQET matrix elements.  This
removes the leading renormalon from $m_c^{\rm pole}$ as well.  Then writing 
$m_c = m_b^{1S} - \delta m_{bc}$, and treating $\delta m_{bc} = m_b-m_c$ as an
independent parameter is practical, as it is well constrained by measured $B\to
X_c\ell\bar\nu$ spectra, and is the dominant source of formally ${\cal
O}(\lambda_1/m_c^2)$ corrections~\cite{Bauer:2004ve}.  (Note that the $B\to
X_c\ell\bar\nu$ data imply that the correlation of these terms with other
contributions is very significant.) Numerically, we use the average of the fit
results in Refs.~\cite{Bauer:2004ve, Amhis:2012bh} and use their difference of
$20\,\MeV$ as a conservative error. For $\lambda_1$ we use $-0.3\,\GeV^2$ as the
central value and vary it by $25\%$, which covers the values obtained in
Refs.~\cite{Bauer:2004ve, Amhis:2012bh} and also the somewhat lower value
implied by the result we use for the leading shape function. The value of
$\lambda_2 = 0.12\,\GeV^2$ is known very well from the $m_{B^*} - m_B$ mass
splitting. We also vary it by $25\%$. The variations for $\lambda_1$ and
$\lambda_2$ can be viewed as an uncertainty estimate to account for the
higher-order perturbative corrections to their OPE coefficients (as well as the
omitted $\lqcd^3/m_b^3$ corrections).

Figure~\ref{fig:spectra} shows the predictions for $\d\Gamma/\d q^2$ (left) and
$\d\Gamma/\d E_\tau$ (right) in the $1S$ mass scheme for the $b$ quark.  The
dotted (green) curves show the free-quark decay result, the dashed (blue) curves
include ${\cal O}(\alpha_s)$ corrections, and the solid (orange) curves include
both $\alpha_s$ and $\lqcd^2/m_b^2$ corrections.  The $\lqcd^2/m_b^2$
corrections are negligible at low values of $\hqs$ and $y$, while their effects
become important for larger values. For $\d\Gamma/\d q^2$, they drive the
spectrum negative near the end point, where the OPE breaks down, as already
discussed above. The peculiar shape of $\d\Gamma/\d E_\tau$ including the ${\cal
O}(\lqcd^2/m_b^2)$ terms is due to the fact that near the end point both the
$\lambda_1$ and $\lambda_2$ terms are large, and the $\lambda_1$ term changes
sign.  For $\d\Gamma/\d E_\tau$ the dot-dashed (dark red) curve combines the
${\cal O}(\alpha_s,\, \lqcd^2/m_b^2)$ corrections with the tree-level leading
shape function result in Eq.~\eqref{eq:dGdySF} (appropriately avoiding any
double counting of $\lqcd^2/m_b^2$ corrections). The theoretical uncertainty of
$\d\Gamma/\d E_\tau$ becomes clearly large for $E_\tau \gtrsim 2.3\,\GeV$, where
the result including shape function effects starts to differ noticeably from the
local OPE result. On the other hand, for $E_\tau \lesssim 2.2\,\GeV$ the local
OPE provides a reliable prediction for the spectrum.

Figure~\ref{fig:spectvar} shows the various sources of uncertainties in the
results in Fig.~\ref{fig:spectra} from varying the parameters as mentioned above
and summarized in Table~\ref{parameters}. The variations from $m_b$ keeping
$\delta m_{bc}$ fixed (solid blue curves) and $\delta m_{bc}$ (dashed light blue
curves) dominate at low and high values, respectively. Varying the
renormalization scale, $\mu$, between $m_b/2$ and $2m_b$ is shown by the solid
green curves, and varying the coefficients of $\lambda_2$ and $\lambda_1$ are
shown by the solid red  and dotted light orange curves, respectively. The
resulting uncertainties in the total rate from each of these parameter
variations are given in Table~\ref{parameters}. For $\d\Gamma/\d E_\tau$ we also
show the relative corrections due to shape function effects (dark red dot-dashed
curve).

Since the largest parts of the uncertainties cancel in the ratio in
Eq.~(\ref{RXc}), yielding a precise SM prediction of the total $B\to
X_c\tau\bar\nu$ rate in Eq.~(\ref{rate}), and the spectra cannot be calculated
reliably point by point near the end points of either $\d\Gamma/\d q^2$ or
$\d\Gamma/\d E_\tau$, in Fig.~\ref{fig:cumulant} we show the integrated rates
above a cut normalized to the total rate,
\beq\label{shapedef}
\widehat\Gamma(q_{\rm cut}^2) = \frac1\Gamma \int_{q_{\rm cut}^2}\!
  {\d\Gamma\over \d q^2}\,, \qquad\!
\widetilde\Gamma(E_{\rm cut}) = \frac1\Gamma \int_{E_{\rm cut}}\!
  {\d\Gamma\over \d E_\tau}\,, \!
\eeq
at different orders in the OPE. The ${\cal O}(\alpha_s)$ corrections have a
negligible effect on these distributions since they do not affect the shape of
the spectra. The yellow band shows the total uncertainty obtained by adding all
uncertainties in quadrature. To obtain the individual uncertainties we apply the
same variations in both numerator and denominator and take the larger of the
up/down variations as the uncertainty. In these normalized event fractions, the
$m_b$ and $\mu$ variations mostly cancel. The total uncertainty essentially
comes from $\delta m_{bc}$ and $\lambda_2$ for $\widehat\Gamma(q^2)$, and from
$\delta m_{bc}$ and $\lambda_{1,2}$ for $\widetilde\Gamma(E_\tau)$. For
$\widehat\Gamma(q_{\rm cut}^2)$ the relative uncertainties in the OPE result
become very large beyond $q_{\rm cut}^2 \gtrsim 10\,\GeV^2$, which is as
expected. For $\widetilde\Gamma(E_{\rm cut})$ the dot-dashed (dark red) curve
shows the effect of including the leading shape function. One can also see here
that the local OPE result starts to become unreliable beyond $E_{\rm cut}
\gtrsim 2.3\,\GeV$.

\section{Summary and Conclusions}

We calculated the inclusive $B\to X_c\tau\nu$ decay distributions in $\tau$
energy and dilepton invariant mass.  Our results for the $\lqcd^2/m_b^2$
corrections to $\d\Gamma/\d q^2$ are new.  We derived predictions for the
spectra using the $1S$ short-distance mass scheme, incorporating the ${\cal
O}(\lqcd^2/m_b^2)$ and ${\cal O}(\alpha_s)$ corrections.  We also studied the
effects of the shape function on the $\tau$ energy end point region. The rates
can be predicted precisely if one makes no cuts in the regions $q^2 \gtrsim
9\,\GeV^2$ and $E_\tau \gtrsim 2.2\,\GeV$.

Recent measurements of the $\bar B \to D\tau\bar\nu$ and $\bar B\to
D^*\tau\bar\nu$ decay rates indicate possible deviations from the standard
model.  The BaBar and Belle measurements of these exclusive modes are consistent
with one another, but are in some tension with LEP measurements of the inclusive
$B\to X_c\tau\bar\nu$ rate.  This makes a new measurement of the inclusive $B\to
X_c\tau\bar\nu$ decay rate particularly timely, especially since no results are
available from the $e^+e^-$ $B$ factories, and measurements may be possible
using the existing data sets.  Given the current tensions, measuring $B\to
X_c\tau\bar\nu$ will also be important with Belle~II data.

Since it might only be possible to measure the inclusive rate in limited regions
of phase space, precise theory predictions for differential distributions are
required, and the calculations presented here should help to improve the
experimental sensitivities.

\acknowledgments

We thank Phillip Urquijo for helpful discussions.
This work was supported in part by the Office of Science, Office of High Energy
Physics, of the U.S.\ Department of Energy under contract DE-AC02-05CH11231 (ZL)
and by the DFG Emmy-Noether Grant No. TA 867/1-1 (FT).

\end{document}